\documentclass[useAMS,usenatbib,usegraphicx]{mn2e}

\usepackage{psfig}

\def\aj{AJ}%
\def\apj{ApJ}%
\def\aap{A\&A}%
\def\mnras{MNRAS}%
\def\pasp{PASP}%
\def\pasj{PASJ}%
\title[The WEBT campaign on PG 1553+113 in 2013]
{The WEBT campaign on the BL Lac object PG 1553+113 in 2013. 
An analysis of the enigmatic synchrotron emission.
}
\author[C. M. Raiteri et al.] 
{C.~M.~Raiteri              $^{ 1}$\thanks{E-mail:raiteri@oato.inaf.it},
A.~Stamerra                 $^{ 1,2}$,
M.~Villata                  $^{ 1}$,
V.~M.~Larionov              $^{ 3,4,5}$,
\newauthor
J.~A.~Acosta-Pulido         $^{ 6,7}$,
M.~J.~Ar\'evalo             $^{ 6,7}$,
A.~A.~Arkharov              $^{ 4}$,
R.~Bachev                   $^{ 8}$,
\newauthor
E.~Ben\'itez                $^{ 9}$,
V.~V.~Bozhilov              $^{10}$,
G.~A.~Borman                $^{11}$,
C.~S.~Buemi                 $^{12}$,
P.~Calcidese                $^{13}$,
\newauthor
M.~I.~Carnerero             $^{ 1,6,7}$,
D.~Carosati                 $^{14,15}$,
R.~A.~Chigladze             $^{16}$,
G.~Damljanovic              $^{17}$,
\newauthor
A.~Di~Paola                 $^{18}$,
V.~T.~Doroshenko            $^{19}$,
N.~V.~Efimova               $^{ 4}$,
Sh.~A.~Ehgamberdiev         $^{20}$,
\newauthor
M.~Giroletti                $^{21}$,
P.~A.~Gonz\'alez-Morales    $^{ 6,7}$,
A.~B.~Grinon-Marin          $^{ 6,7}$,
T.~S.~Grishina              $^{ 3}$,
\newauthor
D.~Hiriart                  $^{22}$,
S.~Ibryamov                 $^{ 8}$,
S.~A.~Klimanov              $^{ 4}$,
E.~N.~Kopatskaya            $^{ 3}$,
\newauthor
O.~M.~Kurtanidze            $^{16,23}$,
S.~O.~Kurtanidze            $^{16}$,
A.~A.~Kurtenkov             $^{10,8}$,
L.~V.~Larionova             $^{ 3}$,
\newauthor
E.~G.~Larionova             $^{ 3}$,
C.~L\'azaro                 $^{ 6,7}$,
A.~L\"ahteenm\"aki          $^{24,25}$,
P.~Leto                     $^{12}$,
G.~Markovic                 $^{26}$,
\newauthor
D.~O.~Mirzaqulov            $^{20}$,
A.~A.~Mokrushina            $^{ 3,4}$,
D.~A.~Morozova              $^{ 3}$,
R.~ M\'ujica                $^{27}$,
\newauthor
S.~V.~Nazarov               $^{11}$,
M.~G.~Nikolashvili          $^{16}$,
J.~M.~Ohlert                $^{28}$,
E.~P.~Ovcharov              $^{10}$,
\newauthor
S.~Paiano                   $^{29,30}$,
A.~Pastor~Yabar             $^{ 6,7}$,
E.~Prandini                 $^{31,32}$,
V.~Ramakrishnan             $^{24}$,
\newauthor
A.~C.~Sadun                 $^{33}$,
E.~Semkov                   $^{ 8}$,
L.~A.~Sigua                 $^{16}$,
A.~Strigachev               $^{ 8}$,
J.~Tammi                    $^{24}$,
\newauthor
M.~Tornikoski               $^{24}$,
C.~Trigilio                 $^{12}$,
Yu.~V.~Troitskaya           $^{ 3}$,
I.~S.~Troitsky              $^{ 3}$,
G.~Umana                    $^{12}$,
\newauthor
S.~Velasco                  $^{ 6,7}$,
and O.~Vince                $^{17}$
}
\begin{document}
\pagerange{\pageref{firstpage}--\pageref{lastpage}} \pubyear{2015}
\maketitle
\label{firstpage}
\begin{abstract}
A multifrequency campaign on the BL Lac object PG 1553+113 was organized by the Whole Earth Blazar Telescope (WEBT) in 2013 April--August, involving 19 optical, two near-IR, and three radio telescopes. The aim was to study the source behaviour at low energies during and around the high-energy observations by the {\em Major Atmospheric Gamma-ray Imaging Cherenkov} ({\em MAGIC}) telescopes in April--July.
We also analyse the UV and X-ray data acquired by the {\em Swift} and {\em XMM-Newton} satellites in the same period.
The WEBT and satellite observations allow us to detail the synchrotron emission bump in the source spectral energy distribution (SED). 
In the optical we found a general bluer-when-brighter trend. 
The X-ray spectrum remained stable during 2013, but a comparison with previous observations suggests that it becomes harder when the X-ray flux increases. 
The long {\em XMM-Newton} exposure reveals a curved X-ray spectrum. 
In the SED, the {\em XMM-Newton} data show a hard near-UV spectrum, while {\em Swift} data display a softer shape that is confirmed by previous {\em HST}-COS and {\em IUE} observations.  
Polynomial fits to the optical--X-ray SED show that the synchrotron peak likely lies in the 4--30 eV energy range, with a general shift towards higher frequencies for increasing X-ray brightness. 
However, the UV and X-ray spectra do not connect smoothly. Possible interpretations include: i) orientation effects, ii) additional absorption, iii) multiple emission components, and iv) a peculiar energy distribution of relativistic electrons. We discuss the first possibility in terms of an inhomogeneous helical jet model.
\end{abstract}

%

\section{Introduction}
BL Lac objects, together with flat-spectrum radio quasars, form the ``blazar" class of active galactic nuclei (AGNs).
Their observed properties, such as strong variability at all frequencies, high and variable polarization, apparent superluminal motion of the radio components, are explained as due to beamed emission from a relativistic jet closely aligned with the line of sight \citep[e.g.][]{urr95}.
The broad-band spectral energy distribution (SED) of these objects shows two bumps: the low-frequency one (from radio to optical/X-rays) is ascribed to synchrotron emission, while the bump at high energies (X-rays to $\gamma$-rays) is likely produced by inverse-Compton scattering of soft photons off the same relativistic electrons responsible for the synchrotron radiation \citep{kon81}\footnote{Alternatively, a hadronic cascade scenario has been proposed \citep[e.g.,][and references therein]{boe13}}.

The source PG 1553+113 is one of the so-called ``high-energy-peaked" BL Lacs (HBL), whose synchrotron peak in the SED typically falls in the UV--X-ray frequency range \citep{falomo1990}. The redshift of PG 1553+113 is still unknown. Many authors tried to set lower and upper limits in an indirect way, from the lack of the host galaxy detection \citep[e.g.][]{treves2007} or from the $\gamma$-ray spectrum \citep[e.g.][]{abdo2010}. In contrast, \citet{danforth2010} derived firm constraints $0.43 < z < 0.58$ from the direct analysis of inter-galactic absorption features visible in the UV spectra acquired by the {\em Hubble Space Telescope}/Cosmic Origins Spectrograph ({\em HST}-COS). 

Very high energies (VHE, $E>100 \rm \, GeV$) observations with {\em MAGIC} in 2005--2009 showed modest flux variability, possibly correlated with the optical one \citep{aleksic2012}. In 2012 the source was found in a flaring state at VHE and X-rays, but not at the GeV energies covered by the {\em Fermi} satellite \citep{aleksic2015}. 
New {\em MAGIC} observations were performed in April, June and July 2013 during a low emission state.
This time, a broad multiwavelength campaign was organized by the Whole Earth Blazar Telescope\footnote{http://www.oato.inaf.it/blazars/webt} (WEBT), an international collaboration born in 1997 to study blazar emission variability \citep[e.g.][]{vil02,rai06b,vil06,boe07,lar08,rai09}. In this paper we present the results of the WEBT monitoring campaign, which was extended to all the 2013 observing season, and explore the synchrotron part of the source spectral energy distribution (SED). To this aim, we complement the radio-to-optical WEBT observations with the UV and X-ray data acquired by the {\em Swift} and {\em XMM-Newton} satellites in the same period. The results of the {\em MAGIC} observations will be analysed in a forthcoming paper (Ahnen et al.\ 2015, in preparation).

The paper is organized as follows. In Sect.\ 2 we present the WEBT light curves in the optical $BVRI$ and near-IR $JHK$ bands as well as radio light curves at three wavelengths. We also derive calibration of an optical photometric sequence in the $BVRI$ bands and perform colour analysis. The {\em Swift} and {\em XMM-Newton} observations are analysed in Sect.\ 3 and 4, respectively, while in Sect.\ 5 the SED shape and variability is discussed with the help of past data by the {\em Swift}, the {\em International Ultraviolet Explorer} (IUE)  and the {\em HST} satellites. Conclusions are drawn in Sect.\ 6.

\section{WEBT data}

Optical observations for the WEBT campaign were performed with 19 telescopes in 17 observatories around the globe. These are listed in Table \ref{obs} with indication of their observing bands.

\begin{table}
\caption{The optical, near-IR and radio observatories participating in the WEBT campaign.}
\label{obs}
\begin{tabular}{llc}
\hline
Observatory & Country & Bands \\
\hline
\multicolumn{3}{c}{\it Optical} \\
Abastumani  & Georgia & $R$ \\
Belogradchik & Bulgaria & $BVRI$ \\
AstroCamp & Spain & $R$ \\
Crimean & Russia & $BVRI$ \\
Michael Adrian & Germany & $BVRI$ \\
Mt. Maidanak & Uzbekistan & $BVRI$ \\
New Mexico Skies & USA & $R$ \\
Plana & Bulgaria & $BVRI$ \\
Rozhen$^1$ & Bulgaria & $BVRI$ \\
San Pedro Martir & Mexico & $R$ \\
Siding Spring & Australia & $R$ \\
Skinakas & Greece & $BVRI$ \\
St. Petersburg & Russia & $BVRI$ \\
Teide & Spain & $R$ \\
Tijarafe & Spain & $R$ \\
Valle d'Aosta & Italy & $BVRI$ \\
ASV$^2$ & Serbia & $BVRI$ \\
\hline
\multicolumn{3}{c}{\it Near-infrared} \\
Campo Imperatore & Italy & $JHK$ \\
Teide & Spain & $JHK$ \\
\hline
\multicolumn{3}{c}{\it Radio} \\
Medicina & Italy & 8 GHz \\
Mets\"ahovi & Finland & 37 GHz \\
Noto & Italy & 43 GHz \\
\hline
$^1$ Three telescopes\\
$^2$ Astronomical Station Vidojevica
\end{tabular}
\end{table}

\begin{figure}
   \centering
   \resizebox{\hsize}{!}{\includegraphics{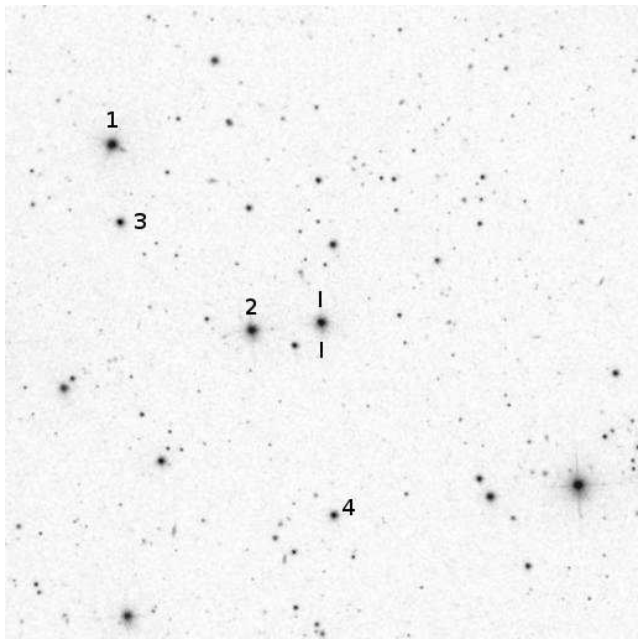}}
   \caption{Finding chart for PG 1553+113 with the stars of the optical photometric sequence labelled from 1 to 4. North is up and East on the left. The field of view is $\sim 7$ arcmin $\times$ 7 arcmin.}
    \label{chart}
    \end{figure}

Calibration of the source magnitude was obtained with respect to the reference stars 1, 2, 3 and 4 shown in Fig.\ \ref{chart}. Photometry of Stars 3 and 4 was derived from the Sloan Digital Sky Survey\footnote{http://www.sdss.org/} (SDSS), using $ugriz$ to $UBVRI$ transformations by \citet{cho08}. The SDSS magnitudes for Stars 1 and 2 are affected by saturation, so we calibrated these stars through differential photometry with respect to Stars 3 and 4 using our best-quality data. The results of our photometry are reported in Table \ref{stars}. 
They are in good agreement with those of \citet{andruchow11} and \citet{doroshenko05,doroshenko14}, while the difference with the values contained in the USNO-A2.0 catalogue is up to several tenths of mag. 

\begin{table*}
\begin{minipage}{100mm}
\caption{Magnitudes (and errors) of the optical photometric sequence shown in Fig.\ \ref{chart} for calibration of the PG 1553+113 magnitude.
The values derived from ground-based observations by the WEBT for the Bessell's filters \citep{bes98} are compared 
to those derived in this paper from {\em Swift}-UVOT and {\em XMM-Newton}-OM observations of the PG 1553+113 field.}
\label{stars}
\begin{tabular}{lcccc}
\hline 
Band & Star 1 & Star 2 & Star 3 & Star 4 \\
\hline
\multicolumn{5}{c}{WEBT} \\
$B$ & 14.503 (0.047) & 14.543 (0.050) & 16.399 (0.037) & 16.555 (0.039) \\
$V$ & 13.832 (0.027) & 13.923 (0.022) & 15.688 (0.017) & 15.771 (0.019) \\
$R$ & 13.465 (0.032) & 13.582 (0.029) & 15.277 (0.026) & 15.317 (0.029) \\
$I$ & 13.080 (0.056) & 13.230 (0.055) & 14.896 (0.051) & 14.893 (0.056) \\
\hline
\multicolumn{5}{c}{{\em Swift}-UVOT} \\
$B$ & 14.497 (0.019) & 14.547 (0.019) & 16.420 (0.051) & 16.580 (0.079) \\
$V$ & 13.885 (0.033) & 13.954 (0.030) & 15.721 (0.034) & 15.837 (0.043) \\
\hline
\multicolumn{5}{c}{{\em XMM-Newton}-OM} \\
$B$ & 14.556 (0.022) & 14.605 (0.022) & 16.501 (0.023) & 16.652 (0.023) \\
$V$ & 13.838 (0.022) & 13.904 (0.022) & 15.676 (0.023) & 15.773 (0.023) \\
\hline
\end{tabular}
\end{minipage}
\end{table*}

\begin{figure*}
   \centering
   \includegraphics[width=12cm]{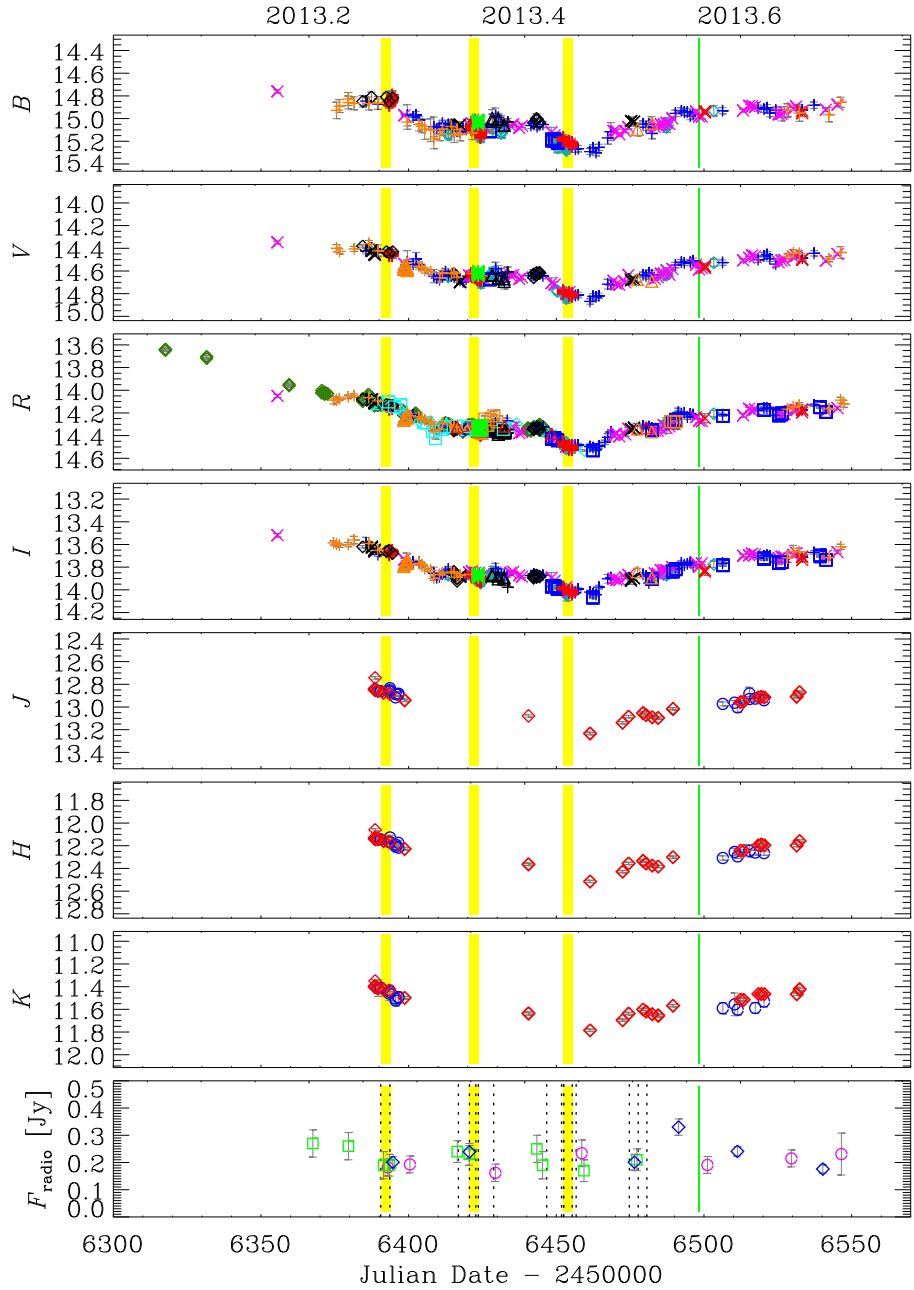}
   \caption{Cleaned light curves of PG 1553+113 built with data from the WEBT collaboration. Different colours and symbols highlight data from different telescopes. 
The yellow and green stripes mark the times of {\em MAGIC} and {\em XMM-Newton} observations, respectively; dotted lines those of {\em Swift} pointings. In the bottom panel purple circles, green squares and blue diamonds refer to flux densities at 43, 37 and 8 GHz, respectively.}
    \label{webt}
    \end{figure*}

In total, we collected 3908 optical data points in the period 2013 January 25 to September 11.
Light curves in $BVRI$ bands were built by carefully assembling the data sets coming from the different telescopes. Cleaned light curves were obtained after correcting for evident outliers, which are specially recognizable through a comparison among the source behaviour in different bands. Moreover, binning was used to reduce the noise of data acquired close in time by the same telescope.
After the cleaning procedure, we were left with 3336 data points.
No shift was given to correct for offsets between different data sets, since in the few cases where an offset was clearly identified, it was always lower than 0.05 mag.
The final light curves are shown in Fig.\ \ref{webt}, where different symbols and colours highlight data from the various telescopes. The epochs of {\em MAGIC}, {\em XMM-Newton} and {\em Swift} observations are also indicated.

Starting from the first data in January, the source brightness smoothly decreased by $\Delta R \sim 0.9$ mag until June, when it reached a minimum; then it started to grow slowly. In the other, less-sampled optical bands the overall variation is about 0.5 mag. No significant flickering (fast continuous variability of the order of few tenths of mag on daily time scales) is visible.
The three {\em MAGIC} observations occurred during the dimming phase, the last one very close to the minimum. In contrast, the {\em XMM-Newton} pointing met the source in the brightening phase.
{\em Swift} observations are more distributed in time.


Near-IR observations were performed at the Campo Imperatore and Teide observatories (see Table \ref{obs}). The data sampling is not as dense as the optical one, but we can recognize a similar trend.

Radio data were acquired at 37~GHz with the 14~m antenna of the Mets\"ahovi Observatory and at 43 and 8~GHz with the 32~m telescopes of the Noto and Medicina stations of the Istituto Nazionale di Astrofisica (INAF), respectively. The source is quite faint at radio wavelengths, its flux oscillating around 0.2 Jy.

\subsection{Colour analysis}

\begin{figure}
   \centering
   \resizebox{\hsize}{!}{\includegraphics{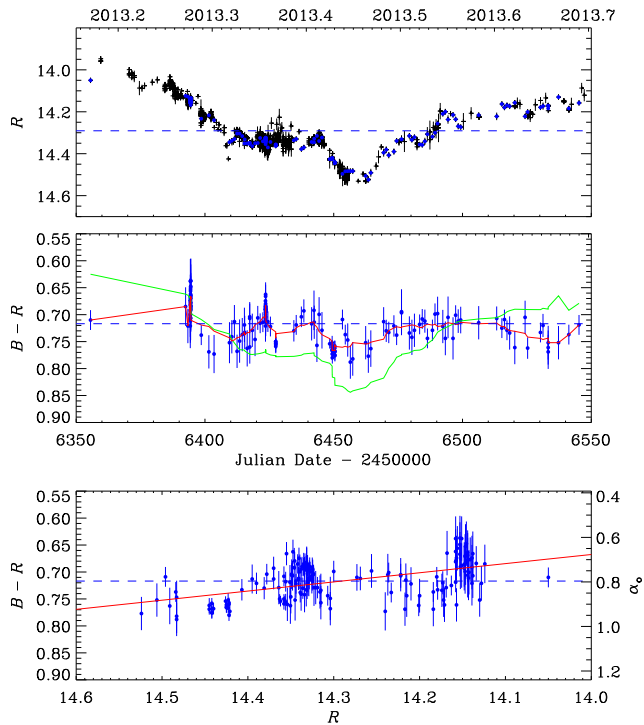}}
   \caption{Top: the $R$-band light curve (black plus signs) where data points used to build the $B-R$ colour indices have been highlighted with blue dots. Middle: $B-R$ versus time; the red line represents the smoothed trend of the colour index; the green line the smoothed $R$-band light curve ($R/2-6.4$). Bottom: $B-R$ and optical spectral index $\alpha_{\rm o}$ versus brightness; the red line represents a linear fit to the data. Horizontal dashed lines mark mean values.}
    \label{colori}
    \end{figure}

Although in the considered period the source was not particularly active, we investigated the presence of possible spectral variations. 
We built $B-R$ colour indices by coupling data acquired from the same telescope within 15 min.
To increase the accuracy of the results, we only considered $R$- and $B$-band data with errors less than 0.02 and 0.04 mag, respectively, obtaining 169 values. The average is $<B-R>$=0.72 and the standard deviation 0.03.
The top panel of Fig.\ \ref{colori} highlights the points in the $R$-band light curve that have been used to build the $B-R$ colour indices. 
These are plotted as a function of time in the middle panel, which also shows a smoothed $B-R$ trend compared to a smoothed $R$-mag trend\footnote{Smoothing was obtained with a boxcar average of width 8, see http://www.exelisvis.com/docs/SMOOTH.html}.
The comparison suggests that most of the time $B-R$ values increase with decreasing flux. This is more evident in the bottom panel,
where the behaviour of the colour index with brightness is displayed: a weak (linear Pearson's correlation coefficient of 0.53 and Spearman's rank correlation coefficient of 0.52 with two-sided significance of its deviation from zero of $3.6 \times 10^{-13}$), but clear, bluer-when-brighter trend is recognizable, which is often found in BL Lacs \citep[see e.g.][]{ikejiri2011,wierzcholska2015}. It is interesting to notice that \citet{wierzcholska2015} did not find a definite trend when analysing the colour index behaviour in 2007--2012, when the source was in a brighter state than in 2013. However, the authors noticed that both bluer-when-brighter and redder-when-brighter trends can be found on shorter time scales. Indeed, a redder-when-brighter tendency is visible in Fig.\ \ref {colori} after $\rm JD=2456500$.

In Fig.\ \ref{colori} we also show the value of the optical spectral index 
$\alpha_{\rm o}=[(B-R)-(A_B-A_R)-0.306]/0.413$, 
assuming that $F_\nu \propto \nu^{-\alpha}$ is the de-reddened flux density\footnote{We adopted effective wavelength and absolute flux values from \citet{bes98}.}. 
The Galactic extinction values in the $B$ and $R$ bands, $A_B$ and $A_R$, are reported in Table \ref{ext}.
We obtain a minimum and a maximum $\alpha_{\rm o}$ of 0.60 and 0.97, with a mean index of $<\alpha_{\rm o}>=0.80$ and a standard deviation of 0.08. 
The spectral trend visible in the bottom panel then indicates that the synchrotron bump in the SED tends to peak very close to the optical band in faint states ($\nu F_\nu \propto \nu^{-\alpha+1} \sim \rm const$), while in brighter states the peak shifts to higher frequencies ($-\alpha+1>0$).

Using the same procedure we estimated the near-infrared spectral index from the $J-K$ colour index as 
$\alpha_{\rm nir}=[(J-K)-(A_J-A_K)-0.987]/0.635$. 
The average value of 45 indices is $<\alpha_{\rm nir}>=0.67$ with a standard deviation of 0.02. In this case there is no recognizable trend with brightness.

We finally calculate the radio--optical spectral index $\alpha_{\rm ro}$ by considering the Medicina flux densities at 8 GHz, $F_8$, and simultaneous $R$-band data,
$\alpha_{\rm ro}=0.210 \, \log F_8+0.084 \, (R-A_R)-0.731$. 
We obtain values in the range 0.30--0.35\footnote{We neglect $k$-correction because we are dealing with an object of uncertain redshift. However, a rough estimate assuming $z=0.5$ and adopting the spectral indices of \citet{fossati1998}, indicates that the $k$-correction should increase the index by only about 0.02.}, in agreement with the typical indices of HBL \citep[e.g.][]{donato2001}.

\begin{table}
\caption{Galactic extinction values (mag) used in this paper for the different photometric bands. 
They are based on the \citet{car89} laws with $R_V=A_V/E(B-V)=3.1$, the standard value for the diffuse interstellar medium, and on $A_B=0.224$ from \citet{sch98}.
In the case of UVOT and OM, the laws were convolved with the filter effective areas and source spectrum.}
\label{ext}
\begin{tabular}{cccc}
\hline
Band & Bessel  & UVOT & OM\\
\hline
$W$2   & -     & 0.457 & 0.484 \\
$M$2   & -     & 0.481 & 0.470 \\
$W$1   & -     & 0.394 & 0.326 \\
$U$    & 0.262 & 0.275 & 0.275 \\
$B$    & 0.224 & 0.229 & 0.230 \\
$V$    & 0.170 & 0.173 & 0.173 \\
$R$    & 0.142 & - & - \\
$I$    & 0.101 & - & - \\
$J$    & 0.049 & - & - \\
$H$    & 0.031 & - & - \\
$K$    & 0.019 & - & - \\
\hline
\end{tabular}
\end{table}

\section{Observations by {\em Swift}}
During the WEBT campaign, the source was observed by the {\em Swift} satellite during 16 epochs.
These are indicated in Fig.\ \ref{webt}. Details on the observations are reported in Table \ref{swift}.

\begin{table}
\caption{{\em Swift} pointings at PG 1553+113: starting times and duration of the XRT and UVOT exposures.}
\label{swift}
\begin{tabular}{llrr}
\hline
ObsID & Start time          & XRT (s) & UVOT (s)\\
\hline
00031368050 & 2013-04-08 01:07:59 &  1793 &   1746\\
00031368051 &2013-04-11 02:57:59 &  2196 &   2222\\
00031368052 &2013-05-04 08:09:59 &  1170 &   1144\\
00031368053 &2013-05-08 02:01:59 &  1954 &   1906\\
00031368054 &2013-05-10 05:37:59 &   859 &    831\\
00031368055 &2013-05-11 02:19:59 &  1307 &   1284\\
00031368056 &2013-05-16 06:58:58 &  1029 &   1003\\
00031368057 &2013-06-03 06:06:59 &  1024 &    997\\
00031368058 &2013-06-08 06:11:59 &   950 &    920\\
00031368059 &2013-06-08 22:33:59 &  1040 &   1011\\
00031368060 &2013-06-09 00:13:58 &   580 &    551\\
00031368061 &2013-06-11 21:13:59 &  1974 &   1918\\
00031368062 &2013-06-13 05:06:59 &  1014 &    988\\
00031368063 &2013-07-01 05:24:59 &  1049 &   1020\\
00031368064 &2013-07-04 05:26:59 &  1030 &   1001\\
00031368065 &2013-07-07 04:07:58 &  1080 &   1053\\
\hline
\end{tabular}
\end{table}

We reduced the data with the {\small HEASOFT} package version 6.15.1.

\subsection{UVOT}
The {\em Swift} satellite carries a 30-cm Ultraviolet/Optical telescope \citep[UVOT;][]{rom05} equipped with $v$, $b$ and $u$ optical filters, and $w1$, $m2$ and $w2$ UV filters.
To process the data, we used the UVOT calibration release 20130118 of the CALDB data base available at NASA's High Energy Astrophysics Science Archive Research Center (HEASARC)\footnote{http://heasarc.nasa.gov/}. For each epoch, possible multiple images in the same filter were first summed with the task {\tt uvotimsum} and then aperture photometry was performed with {\tt uvotsource}. We extracted source counts from a circular region with 5 arcsec radius centred on the source and background counts from a circle with 20 arcsec radius in a source-free field region.

\begin{figure}
   \centering
   \resizebox{\hsize}{!}{\includegraphics{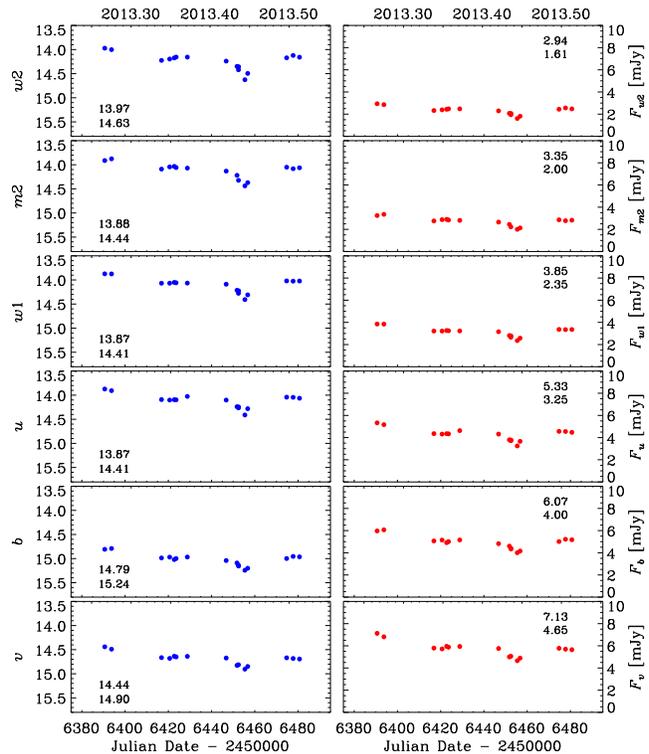}}
   \caption{UV--optical light curves of PG 1553+113 obtained from {\em Swift}-UVOT data. Observed magnitudes are plotted on the left; maximum and minimum values are indicated in the lower left. The flux densities on the right were obtained after recalibration and correction for the Galactic extinction as explained in the text; maximum and minimum values are printed in the upper right.}
    \label{uvot}
    \end{figure}

The UVOT light curves are shown in Fig.\ \ref{uvot}. On the left we plotted the observed magnitudes, while on the right we show flux densities after recalibration and correction for the Galactic extinction. Recalibration is necessary when a source has a spectral shape that differs from that of the objects used to calibrate the instrument.
The average $b-v$ of PG 1553+133 is 0.33, which is out of the colour range inside which the available UV count rate to flux conversion factors (cfs) are valid \citep{poo08,bre11}.
We thus interpolated the source spectrum with a log-linear fit and convolved it with the filter effective areas to calculate new factors \citep[see e.g.][]{rai10}. 
The procedure was iterated to check the stability of the results.
The ratio between these new cfs and the old cfs by \citet{bre11} is 
0.997, 1.001, 1.015, 1.064, 0.995 and 1.014 in the
$v$, $b$, $u$, $w1$, $m2$ and $w2$ bands,
respectively. The most important effect of the recalibration is thus to enhance the source flux density in the $w1$ band by about 6\%. 

Galactic extinction in each UVOT band was similarly estimated by convolving the \citet{car89} laws with the filter effective areas and source spectrum, adopting $R_V=A_V/E(B-V)=3.1$ (the standard value for the diffuse interstellar medium), and starting from an extinction of 0.224 mag in the Johnson's $B$ band \citep{sch98}. The results after iteration are reported in Table \ref{ext}.
A discussion on the uncertainties affecting estimates of Galactic extinction, which are larger in the UV, can be found in \citet{fit07}. In the case of PG 1553+113, which has a low $E(B-V)=0.054$ (see Table \ref{ext}), we would expect an error of 0.02--0.04 mag on the UV de-reddened continuum due to uncertainties in the extinction alone.

The UVOT light curves plotted in Fig.\ \ref{uvot} confirm the trend shown by the ground-based optical (and near-IR) curves in Fig.\ \ref{webt}.
They also suggest an increasing variability amplitude with frequency, as usually observed in BL Lac objects \citep[e.g.][]{rai10}.

We also derived UVOT $v$ and $b$ magnitudes for the reference stars and transformed them into Johnson's magnitudes through the colour corrections given by \citet{poo08}. The resulting $V$ and $B$ magnitudes, which differ from the $v$ and $b$ values by 0.009--0.017 mag, are reported in Table \ref{stars}. They are in excellent agreement with the magnitudes derived from the WEBT data in the $B$ band, while the UVOT data are a bit fainter in the $V$ band.


\subsection{XRT}
The {\em Swift} X-ray Telescope \citep[XRT;][]{bur05} is a CCD imaging spectrometer. We reduced the data with the CALDB calibration files updated 20140120. We run the task {\tt xrtpipeline} with standard screening criteria on the observations performed in pointing mode. We found four observations in Photon Counting (PC) mode and 16 observations in Windowed Timing (WT) mode, three of which were discarded because they had exposure times less than 30 s. The count rate ranges from 1.3 to 3.0 counts $\rm s^{-1}$ hence the four PC exposures need correction for pile-up. 
We analysed the PC images with the task {\tt ximage} and modelled the wings of the source point spread function (PSF) with the King's function representing the expected PSF of XRT \citep{mor05}. The extrapolation of the fit to the inner region allowed us to define the radius within which pile-up is important. 
This turned out to be 10 arcsec. We thus extracted source counts from an annular region with inner and outer radius of 10 and 75 arcsec, respectively, while we took the background from an annulus between 100 and 150 arcsec.
As for the observations performed in WT mode, we extracted the source counts in a circular region with 70 arcsec radius, and the background counts from a region of the same size shifted along the window, away from the source PSF.
The task {\tt xrtmkarf} was used to create ancillary response files to correct for
hot columns, bad pixels, and the loss of counts caused by using an annular extraction region in the pile-up case.

We grouped each spectrum with the corresponding background, redistribution matrix (rmf), and ancillary (arf) files with the task {\tt grppha}, setting a binning of at least 25 counts for each spectral channel in order to use the $\chi^2$ statistic.
The spectra were analysed with {\small XSPEC} version 12.8.1.
We adopted a Galactic absorption of $N_{\rm H} =0.372 \times 10^{21} \rm \, cm^{-2}$ from the LAB survey \citep{kal05} and the \citet{wil00} elemental abundances.
The spectra were fitted with both an absorbed power law $N(E)=N_0 \, E^{-\Gamma}$, where $N_0$ represents the number of photons $\rm keV^{-1} \, cm^{-2} \, s^{-1}$ at 1 keV, and absorbed curved models.
Application of a broken power law model 
$$N(E)= \left\{
\begin{array}{l l}
N_0 \, E^{-\Gamma_1} & \quad E<E_{\rm b} \\
N_0 \, E_{\rm b}^{\Gamma_2-\Gamma_1} \, E^{-\Gamma_2} & \quad E \ge E_{\rm b}
\end{array} \right.
$$
led to large errors on the parameters or to multiple solutions.
We also tried the log-parabola model \citep{lan86,mas04}
$$N(E)=N_0 \, (E/E_{\rm s})^{-\alpha-\beta \log (E/E_{\rm s})},$$
(where $E_{\rm s}$ is a scale parameter that we fixed equal to 1 keV) which has largely been used to fit the X-ray spectrum of this source (see Sect.\ \ref{discussion}).

Details of the power-law fits are given in Table \ref{xrt}, where Column 1 gives the observation ID, Column 2 the XRT observing mode, Column 3 the half-exposure JD, Column 4 the 1 keV flux density corrected for the Galactic extinction, Column 5 the power-law photon index $\Gamma$, Column 6 the reduced $\chi^2$, and Column 7 the number of degrees of freedom. 
The average photon index is $\Gamma=2.14$, with a standard deviation of 0.06, and there is no significant correlation with brightness. Indeed, the linear Pearson correlation coefficient is only 0.46 and the variance is less than the mean square uncertainty, which means that possible variability is hidden by errors.
Table \ref{logpar} gives the same information as Table \ref{xrt} for the log-parabolic fits, where the photon index $\Gamma$ is replaced by the two parameters $\alpha$ and $\beta$. In the last Column we also report the $F$-test probability $P_F$ (see below).

Most of the power-law fits are statistically good ($\chi^2_{\rm red} \sim 1$), with a few exceptions. 
The most critical case is the observation at $\rm JD=2456422.74$, for which the fit cannot be improved by the curved model. 
In contrast, the data acquired on $\rm JD=2456428.80$ seem to be better reproduced by a log-parabola.
To verify when the log-parabola offers a statistically better fit to the data than the power-law model, we applied the $F$ statistic, which can be used to test for an additional term in a model \citep[e.g.,][]{protassov2002,orlandini2012}. A low $F$-test probability $P_F$ (e.g.\ below $\sim 0.01$) means that 
the simpler model is unlikely to be correct.
The $P_F$ in Table \ref{logpar} covers a wide range of values,
favouring the log-parabola at some epochs and disfavouring it in the others. The curvature parameter $\beta$ corresponding to the cases with $P_F < 0.01$ goes from 0.27 to 0.53, indicating noticeable curvature, as expected. 
When instead $\beta$ approaches zero, $P_F$ approaches one, or is not even computable.
However, the large uncertainties on $\beta$ as well as its strong erratic variations prevent a reliable characterization of the X-ray spectral curvature with XRT.

\begin{figure}
   \centering
   \resizebox{\hsize}{!}{\includegraphics{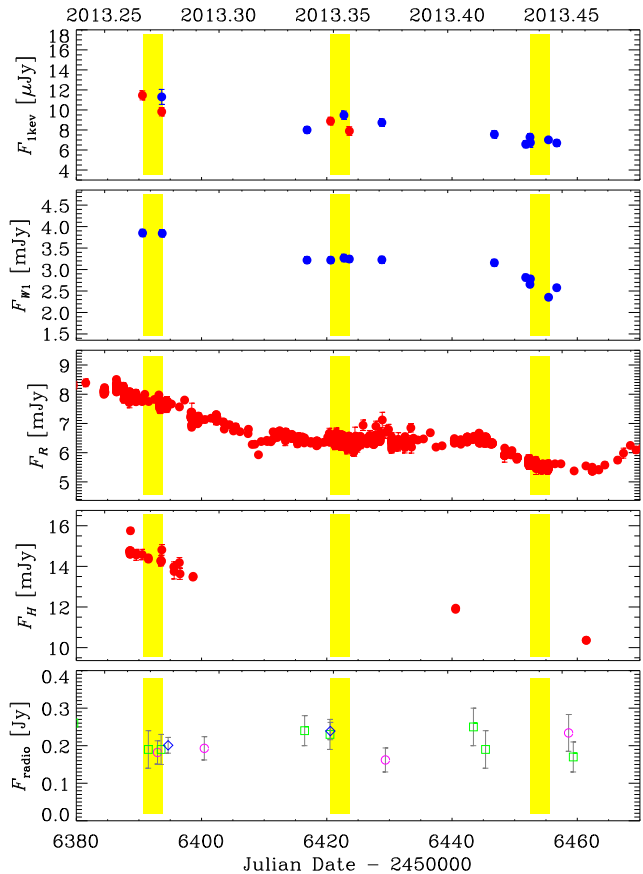}}
   \caption{Multiwavelength light curves of PG 1553+113 around the period of the {\em MAGIC} pointings (yellow stripes). In the top panel the de-absorbed flux density at 1 keV is given, separating the data obtained in PC (red dots) from WT (blue dots) mode. Flux densities in the $w1$, $R$ and $H$ bands are corrected for Galactic extinction as explained in the text. Different colours and symbols in the bottom panel, showing the radio flux densities, have the same meaning as in Fig.\ \ref{webt}.}
    \label{mw}
    \end{figure}

In Fig.\ \ref{mw} the de-absorbed X-ray light curve (obtained through power law fits) is compared with those at lower frequencies, showing the same decreasing trend of the near-IR to UV light curves.

\begin{table*}
\caption{Results of fitting the {\em Swift}-XRT spectra with an absorbed power-law model with photon index $\Gamma$ and fixed Galactic absorption $N_{\rm H} =0.372 \times 10^{21} \rm \, cm^{-2}$.}
\label{xrt}
\begin{tabular}{ccccccr}
\hline
ID & Mode  & JD & $F_{\rm 1keV}$ [$\mu$Jy] & $\Gamma$ & $\chi^2_{\rm red}$ & d.o.f.\\
\hline
\multicolumn{7}{c}{2013}\\
00031368050 & PC & 2456390.55759 &   11.46   (11.01--11.91)  & 2.14  (2.09--2.20) &    0.95 &     77\\
00031368051 & PC & 2456393.63631 &    9.83   (9.40--10.25)   & 2.12  (2.05--2.18) &    1.17 &     71\\
00031368051 & WT & 2456393.63631 &   11.30   (10.56--12.05)  & 2.19  (2.09--2.29) &    1.07 &     46\\
00031368052 & WT & 2456416.84703 &    8.01    (7.67--8.34)   & 2.15  (2.09--2.22) &    0.99 &     92 \\
00031368053 & PC & 2456420.59602 &    8.89    (8.51--9.26)   & 2.15  (2.09--2.20) &    1.07 &     72 \\
00031368054 & WT & 2456422.73968 &    9.48    (9.06--9.89)   & 2.14  (2.08--2.21) &    1.40 &     81 \\
00031368055 & PC & 2456423.60478 &    7.89    (7.46--8.32)   & 2.11  (2.03--2.19) &    0.97 &     45 \\
00031368056 & WT & 2456428.79691 &    8.74    (8.36--9.11)   & 2.19  (2.13--2.25) &    1.21 &     86 \\  
00031368057 & WT & 2456446.76078 &    7.56    (7.20--7.93)   & 2.09  (2.02--2.16) &    1.09 &     72 \\
00031368058 & WT & 2456451.76382 &    6.58    (6.23--6.94)   & 2.08  (2.00--2.16) &    1.10 &     63 \\
00031368059 & WT & 2456452.44628 &    7.29    (6.95--7.62)   & 2.24  (2.17--2.32) &    0.85 &     74 \\
00031368060 & WT & 2456452.51305 &    6.74    (6.28--7.20)   & 2.01  (1.92--2.11) &    0.93 &     41 \\
00031368061 & WT & 2456455.39613 &    7.01    (6.77--7.26)   & 2.17  (2.12--2.22) &    0.98 &    131\\
00031368062 & WT & 2456456.71905 &    6.70    (6.37--7.03)   & 2.11  (2.04--2.19) &    0.93 &     73\\
00031368063 & WT & 2456474.73176 &    9.38    (8.99--9.78)   & 2.15  (2.09--2.21) &    1.17 &     91\\
00031368064 & WT & 2456477.73303 &   10.82   (10.37--11.26)  & 2.20  (2.14--2.26) &    0.92 &     94\\
00031368065 & WT & 2456480.67845 &   10.35   (9.95--10.74)   & 2.21  (2.16--2.27) &    1.01 &    102\\
\hline
\multicolumn{7}{c}{2009}\\
00031368009 & PC & 2455083.53235 &    2.63   (2.54--2.72)    & 2.37  (2.33--2.42) &    1.11 &     92\\   
00031368010 & PC & 2455108.09667 &    2.18   (2.08--2.28)    & 2.43  (2.36--2.51) &    1.06 &     54\\
\hline
\multicolumn{7}{c}{2012}\\
00031368035 & WT & 2456046.57654 &   17.12  (16.61--17.63)   & 2.16  (2.12--2.20) &    1.15 &     144\\
\hline
\end{tabular}
\end{table*}

\begin{table*}
\caption{Results of fitting the XRT spectra with an absorbed log-parabola model $N(E)=N_0 \, (E/E_{\rm s})^{-\alpha-\beta \log (E/E_{\rm s})}$ with $E_{\rm s}=1 \, \rm keV$ and fixed Galactic absorption $N_{\rm H} =0.372 \times 10^{21} \rm \, cm^{-2}$.}
\label{logpar}
\begin{tabular}{cccccccrc}
\hline
ID & Mode  & JD & $F_{\rm 1keV}$ [$\mu$Jy] & $\alpha$ & $\beta$ & $\chi^2_{\rm red}$ & d.o.f. & $P_F$\\
\hline
\multicolumn{7}{c}{2013}\\
00031368050 & PC & 2456390.55759 &   12.26   (11.66--12.86)  & 2.06  (1.99--2.13) &    0.37  (0.20--0.56) &    0.80 &     76 & $2 \times 10^{-4}$\\
00031368051 & PC & 2456393.63631 &   10.19   (9.66--10.72)   & 2.06  (1.98--2.14) &    0.22  (0.04--0.42) &    1.14 &     70 & 0.09\\
00031368051 & WT & 2456393.63631 &   11.44   (10.54--12.31)  & 2.16  (2.02--2.29) &    0.09  ($-0.18$--0.40)&    1.09 &   45 & 0.69\\
00031368052 & WT & 2456416.84703 &    8.17    (7.76--8.58)   & 2.12  (2.03--2.20) &    0.14  ($-0.06$--0.35)&    0.98 &   91 & 0.17\\
00031368053 & PC & 2456420.59602 &    9.77   (9.26--10.27)   & 2.03  (1.94--2.10) &    0.52  (0.32--0.73)&    0.79 &      71 & $2 \times 10^{-6}$\\
00031368054 & WT & 2456422.73968 &    9.56   (9.07--10.07)   & 2.13  (2.04--2.21) &    0.06  ($-0.13$--0.26)&    1.42 &   80 & -\\
00031368055 & PC & 2456423.60478 &    8.26    (7.71--8.81)   & 2.05  (1.95--2.15) &    0.28  (0.03--0.54) &    0.92 &     44 & 0.07\\
00031368056 & WT & 2456428.79691 &    9.41    (8.94--9.88)   & 2.06  (1.97--2.15) &    0.53  (0.32--0.75) &    1.01 &     85 & $6 \times 10^{-5}$\\  
00031368057 & WT & 2456446.76078 &    7.96    (7.51--8.41)   & 1.99  (1.88--2.08) &    0.37  (0.15--0.62) &    0.99 &     71 & $5 \times 10^{-3}$\\
00031368058 & WT & 2456451.76382 &    6.96    (6.53--7.36)   & 1.97  (1.85--2.08) &    0.42  (0.15--0.72) &    1.01 &     62 & 0.01\\
00031368059 & WT & 2456452.44628 &    7.42    (6.96--7.83)   & 2.22  (2.13--2.30) &    0.12  ($-0.10$--0.35)&    0.85 &   73 & 0.32 \\
00031368060 & WT & 2456452.51305 &    7.23    (6.66--7.81)   & 1.88  (1.74--2.02) &    0.53  (0.19--0.90) &    0.78 &     40 & $5 \times 10^{-3}$\\
00031368061 & WT & 2456455.39613 &    7.31    (7.00--7.61)   & 2.10  (2.03--2.17) &    0.28  (0.12--0.45) &    0.93 &    130 & $5 \times 10^{-3}$\\
00031368062 & WT & 2456456.71905 &    6.83    (6.44--7.22)   & 2.08  (1.97--2.17) &    0.14  ($-0.08$--0.37)&    0.92 &   72 & 0.18\\
00031368063 & WT & 2456474.73176 &    9.53   (9.05--10.01)   & 2.12  (2.04--2.20) &    0.11  ($-0.09$--0.31)&    1.18 &   90 & 0.63\\
00031368064 & WT & 2456477.73303 &   10.82   (10.29--11.35)  & 2.20  (2.12--2.27) &    0.00  ($-0.18$--0.19)&    0.93 &   93 & -\\
00031368065 & WT & 2456480.67845 &   10.70   (10.22--11.19)  & 2.15  (2.08--2.23) &    0.23  (0.06--0.42) &    0.97 &    101 & 0.02\\
\hline
\multicolumn{7}{c}{2009}\\
00031368009 & PC & 2455083.53235 &    2.82  (2.70--2.95) &  2.32 (2.26--2.38)  &  0.39 (0.22--0.57)    & 0.95 & 91 & $10^{-4}$\\   
00031368010 & PC & 2455108.09667 &    2.25  (2.12--2.39) &  2.42 (2.34--2.50)  &  0.17 ($-0.04$--0.41) & 1.04 & 53 & 0.16\\
\hline
\multicolumn{7}{c}{2012}\\
00031368035 & WT & 2456046.57654 &  17.91 (17.27--18.56) & 2.10 (2.04--2.15)   & 0.27 (0.14--0.41)  & 1.07   & 143 & $8 \times 10^{-4}$\\
\hline
\end{tabular}
\end{table*}

\section{Observations by {\em XMM-Newton}}
A long calibration pointing at PG 1553+113 was performed by the {\em XMM-Newton} satellite on 2013 July 24--25 ($\rm JD=2456498.13560$--2456498.52237).
The total exposure time was 34500 s.
We processed the data with the Science Analysis Software\footnote{http://xmm.esac.esa.int/sas/current/documentation/} ({\small SAS}) package version 13.5.0.

\subsection{OM}
The Optical Monitor \citep{mas01} onboard {\em XMM-Newton} performed nine exposures, six of which with the optical and UV wide-band filters, two with the UV grism (grism1), and one with the optical grism (grism2).
We ran the pipeline {\tt omichain} and analysed the $V$, $B$, $W1$, $M2$ and $W2$ images with the {\tt omsource} task.
The resulting source magnitudes, already adjusted to the Johnson's system, are reported in Table \ref{om}. 
The uncertainties take into account both the measurement and the calibration errors.
The measurement error is very small, of the order of 0.001--0.003 mag,
in all bands but $W2$, where it is 0.018 mag. The errors in the calibration are estimated to be 2\% in $V$ and $B$ bands and 10\% in $U$ band\footnote{http://xmm2.esac.esa.int/docs/documents/CAL-TN-0019.pdf}. We assumed a 10\% error also in the UV bands.
We notice that the OM $V$ and $B$ magnitudes fairly match the simultaneous values by the WEBT shown in Fig.\ \ref{webt}.

\begin{table}
\caption{Results of the {\em XMM-Newton}-OM photometry on PG 1553+113.}
\label{om}
\begin{tabular}{lcccc}
\hline
Filter & Exp [s] & Mag    & $F_\lambda^1$ & $F_\nu^2$\\
\hline
$V$    & 1700     & 14.561 (0.022) & 5.72 & 6.60\\ 
$B$    & 2001     & 14.986 (0.022) & 6.65 & 5.55\\ 
$U$    & 2000     & 14.09 (0.11)   & 9.05 & 4.60\\ 
$W1$   & 3999     & 13.84 (0.11)   & 10.6 & 4.04\\ 
$M2$   & 3999     & 13.79 (0.11)   & 13.6 & 3.74\\ 
$W2$   & 4000     & 13.87 (0.11)   & 14.4 & 3.36\\ 
\hline
\multicolumn{5}{l}{$^1$ Observed flux densities in units $10^{-15} \rm \, erg \, cm^{-2} \, s^{-1} \, \AA^{-1}$}\\
\multicolumn{5}{l}{$^2$ De-absorbed flux densities in mJy}\\
\end{tabular}
\end{table}

We also extracted the $V$ and $B$ magnitudes of the reference stars, to compare their space photometry with the ground-based one.
As can be seen in Table \ref{stars}, the OM values in $V$ band are in excellent agreement with those obtained with the WEBT data, with deviations of less than 0.02 mag. The difference in $B$ band is more pronounced, up to $\sim 0.1$ mag.

The OM grism1 and grism2 provide low-resolution spectra in the range 1800 to 3600 \AA\ and 3000 to 6000 \AA, respectively. Exposure times were 4400 and 4260 s for the two grism1 observations and 4000 s for the grism2. We reduced these data with the task {\tt omgchain} and checked the results with {\tt omgsource}.
The resulting UV spectra are very noisy and no reliable feature can be recognized. 
The optical spectrum is contaminated by zero-order features. 

\begin{figure}
   \centering
   \resizebox{\hsize}{!}{\includegraphics{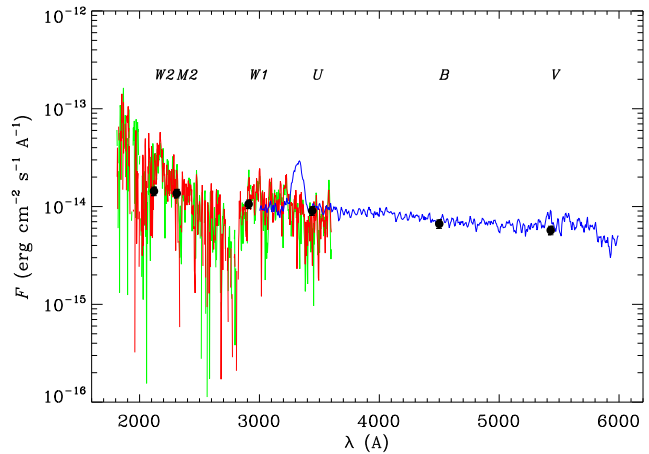}}
   \caption{{\em XMM-Newton}-OM grism spectra and broad-band photometry. The green and red lines represent the grism1 (UV) spectra, while the blue line shows the grism2 (optical) spectrum. Black dots mark the flux densities in the $V$, $B$, $U$, $W1$, $M2$ and $W2$ filters. A conservative 10\% error is assumed on the flux.}
    \label{spectra}
    \end{figure}

The grism spectra are shown in Fig.\ \ref{spectra}, where they are compared with the OM broad-band photometry. Flux densities in the $V$, $B$, $U$, $W1$, $M2$ and $W2$ bands were obtained by multiplying the count rates for the conversion factors derived from observations of white dwarf standard stars\footnote{http://xmm.esac.esa.int/sas/8.0.0/watchout/. We notice that when using conversion factors based on the Vega spectrum, a dip in $U$ band appears.}. They are also reported in Table \ref{om}. 
We assumed a conservative 10\% error on these values. 
There is a fair agreement between the results of the OM spectroscopy and  photometry.

In Table \ref{om} we also give de-absorbed flux densities in mJy.
The values of Galactic extinction for the OM filters calculated by integrating the source spectrum with the \citet{car89} laws and filter effective areas are reported in Table \ref{ext}.

\subsection{EPIC}
The European Photon Imaging Camera (EPIC) onboard {\em XMM-Newton} carries three detectors: MOS1, MOS2 \citep{tur01} and pn \citep{str01}.
Because of the expected high source brightness, exposures were performed in small window mode with medium filter.
We processed the data with the {\tt emproc} and {\tt epproc} tasks of the {\small SAS} package.
Possible high-background periods were removed by asking that the count rate of high-energy events ($> 10 \, \rm keV$) was less than 0.35 and 0.40 cts $\rm s^{-1}$ on the MOS and pn detectors, respectively. 
Out-of-time events are visible in the pn image, but because of the small window mode, their fraction is $\sim 1\%$ only and a correction is not necessary.
We extracted source counts from a circular region with 40 arcsec radius.
In the case of MOS detectors, background was extracted from a circle of 80 arcsec radius on an external CCD, while in the case of pn it was taken from a circle of 40 arcsec radius on the same CCD, as far as possible from the source.
We selected the best calibrated single and double events only (PATTERN$<$=4), and rejected events next to either the edges of the CCDs or bad pixels (FLAG==0).
The absence of pile-up was verified with the {\tt epatplot} task.

As in the XRT data analysis, we grouped each spectrum with the corresponding background, redistribution matrix (rmf), and ancillary (arf) files with the task {\tt grppha}, setting a binning of at least 25 counts for each spectral channel in order to use the $\chi^2$ statistic.
The three spectra were analysed with {\small  XSPEC} version 12.8.1.
We adopted a Galactic absorption of $N_{\rm H} =0.372 \times 10^{21} \rm \, cm^{-2}$ from the LAB survey \citep{kal05} and the \citet{wil00} elemental abundances.
We fitted the three EPIC spectra together. An absorbed power law gives a poor fit to the spectrum, even when letting the absorption free.
The X-ray spectra are better reproduced by absorbed curved models such as a broken power law or a log-parabola (see the previous section).
Table \ref{epic} shows the results of the spectral fitting with the different models. 
Both curved models give a reasonably good fit according to the $\chi^2_{\rm red}$ value. 
Unfortunately, this is not a case where we can use the F-statistic to see which model is better \citep{protassov2002,orlandini2012}, so we consider both as reliable.

\begin{table*}
\caption{Results of fitting the {\em XMM-Newton}-EPIC spectra with different models: power law (pow), broken power law (bknpow) and log-parabola (logpar). For the broken power law and log-parabola models the absorption was fixed to the Galactic value $N_{\rm H} =0.372 \times 10^{21} \rm \, cm^{-2}$, while for the single power law models both free and fixed absorption were tried.} 
\label{epic}
\begin{tabular}{cccccccr}
\hline
Model  & $F_{\rm 1keV}$   [$\mu$Jy] & $N_{\rm H}$ [$10^{21} \rm \, cm^{-2}$] & $\Gamma$  & & $\chi^2_{\rm red}$ & d.o.f.\\                    
pow    & $10.48 \pm 0.02$ & 0.372       & $2.319 \pm 0.002$ & & 1.56               & 3270  \\
\hline
Model  & $F_{\rm 1keV}$  [$\mu$Jy] & $N_{\rm H}$ [$10^{21} \rm \, cm^{-2}$]  & $\Gamma$  & & $\chi^2_{\rm red}$ & d.o.f.\\                    
pow    & $11.15 \pm 0.04$ & $0.550 \pm 0.009$ & $2.395 \pm 0.004$ & & 1.19               & 3269  \\
\hline
Model  & $F_{\rm 1keV}$ [$\mu$Jy] &$\Gamma_1$ & $\Gamma_2$& $E_b$ [keV] & $\chi^2_{\rm red}$ & d.o.f.\\
bknpow & $10.75 \pm 0.03$ & $2.246 \pm 0.005$ & $2.464 \pm 0.010$ & $1.62 \pm 0.06$ & 1.08 & 3268\\
\hline
Model  & $F_{\rm 1keV}$ [$\mu$Jy] & $\alpha$  & $\beta$ & $E_s$ [keV] & $\chi^2_{\rm red}$ & d.o.f.\\
logpar & $10.83 \pm 0.03$ & $2.283\pm 0.003$ & $0.166 \pm 0.007$ & 1 &1.09 & 3269\\
\hline
\end{tabular}
\end{table*}

\begin{figure}
   \centering
   \resizebox{\hsize}{!}{\includegraphics[angle=-90]{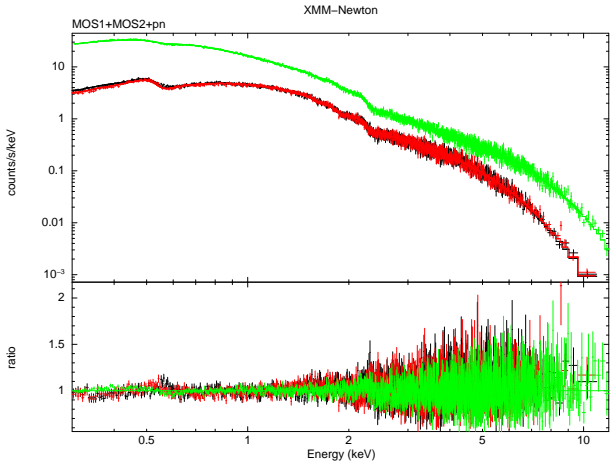}}
   \caption{X-ray spectra of PG 1553+113 acquired by the EPIC instrument onboard {\em XMM-Newton} on 2013 July 24--25. Black, red and green symbols represent MOS1, MOS2 and pn data, respectively. The folded model, an absorbed log-parabola with $N_{\rm H} =0.372 \times 10^{21} \rm \, cm^{-2}$, is shown by solid lines of the same colour. The ratio between the data and the folded model is plotted in the bottom panel.}
    \label{xmmspec}
    \end{figure}

In the parabolic case, the observed fluxes in the 0.3--10, 0.3--2 and 2--10 keV energy ranges are $(6.64 \pm 0.02)$, $(4.27 \pm 0.01)$ and $(2.37 \pm 0.01) \times 10^{-11} \rm \, erg \, cm^{-2} \, s^{-1}$, respectively. The corresponding de-absorbed values are 7.62, 5.25 and $2.37\times 10^{-11} \rm \, erg \, cm^{-2} \, s^{-1}$.
The three EPIC spectra with folded model and data-to-model ratio are shown in Fig.\ \ref{xmmspec}.

We finally extracted source X-ray light curves with different time binning, but they did not show significant variability.

\section{Spectral energy distribution}

In the following, we present SEDs of the source from the near-infrared to the X-ray energies, i.e., the frequency range where the synchrotron bump achieves its maximum in the high-energy peaked BL Lac objects such as PG 1553+113.
For the UV data of both UVOT and OM we assumed a 10\% error on the flux. The errors on the WEBT data also include uncertainties on the calibration. The {\em Swift}-XRT spectra are here modelled with power laws.

\begin{figure}
   \centering
   \resizebox{\hsize}{!}{\includegraphics{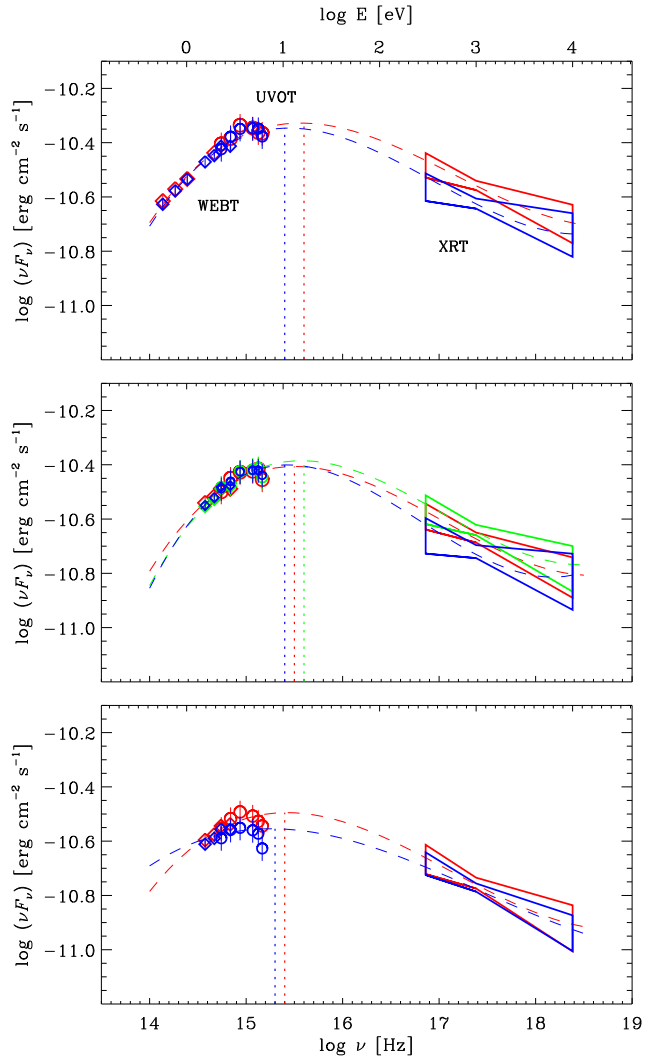}}
   \caption{Spectral energy distribution of PG 1553+113 during the first (top), second (middle) and third (bottom) {\em MAGIC} observations. 
Data from the WEBT are displayed as diamonds, while those from {\em Swift}-UVOT as circles. The errors on the WEBT data take into account also the uncertainties on the calibration. The dashed lines represent log-cubic fits, which suggest a general shift of the synchrotron peak (marked by vertical dotted lines) towards higher frequencies when the source brightness increases.}
    \label{sed_swift}
    \end{figure}

At the epoch of the first {\em MAGIC} pointing, from $\rm JD=2456390.6326$ to $\rm JD=2456393.7306$, two Swift observations were performed: the first one just before the start and the second by the end of the {\em MAGIC} pointing.
Figure \ref{sed_swift} shows the corresponding SEDs of the source; X-ray and UV data from {\em Swift} are complemented by simultaneous optical and near-IR data acquired by the WEBT. 
In particular, in the first SED (red) the $J$, $H$ and $K$ data are from Campo Imperatore, while the $R$-band datum is the average of eight cleaned data from Abastumani. Both a PC-mode and a WT-mode spectra were available for the second SED (blue);  we chose to plot the results of the PC mode because of the larger number of counts. Near-IR points for this epoch represent averages of simultaneous WEBT data from Campo Imperatore and Teide; optical $B$, $V$, $R$ and $I$ data are mean values from simultaneous observations at Rozhen and a dense monitoring at Mt.\ Maidanak.
We notice that the WEBT points nicely continue the spectral shape traced by the optical--UV data from {\em Swift}-UVOT and that the $V$-band points from UVOT and from the WEBT available in the second epoch are in excellent agreement.
The only critical point is in the $B$ band, where the WEBT fluxes are a bit fainter than the UVOT values. This problem has already been found for 3C~454.3 \citep{rai11}. 

In the period of the second {\em MAGIC} pointing ($\rm JD=2456420.5826$--2456423.6299) there were three {\em Swift} observations. 
WEBT $BVRI$ data corresponding to the first epoch are from the Crimean Observatory, complemented by observations at the Abastumani Observatory in the $R$ band.
WEBT data for the second pointing were provided by the Crimean, Mt.\ Maidanak and New Mexico Skies observatories, while for the third pointing WEBT data were acquired at the Valle d'Aosta Observatory.

The third {\em MAGIC} pointing, from $\rm JD=2456452.4486$ to $\rm JD=2456455.5083$, saw again three {\em Swift} pointings. In the figure, we only plot the first and last ones, since the second is very close in time to the first but lacks observations in the $m2$ and $v$ bands.
WEBT data for these two epochs are from the Mt.\ Maidanak Observatory, complemented in the last epoch by observations from the Crimean Observatory.
No near-IR data are available during the second and third {\em MAGIC} observations.

A log-parabolic model \citep{lan86,mas04} applied to the near-IR--X-ray SEDs gave a poor fit to the X-ray data ($\chi^2_{\rm red}=7$--14). Hence we fitted all SEDs with a log-cubic model ($\chi^2_{\rm red}=2$--7), which suggests synchrotron peak frequencies in the range  $\log \nu_{\rm p} = 15.3$--15.6, i.e.\ $\nu_{\rm p}=8$--16 eV, and a general increase with the source brightness.
An increase of the synchrotron peak frequency with flux has already been found for various other blazars, notably for another high-energy peaked BL Lac object: Mkn 501 \citep{pia99}.
Note however that the fit cannot account for the downturn in the UV band.

\begin{figure}
   \centering
   \resizebox{\hsize}{!}{\includegraphics{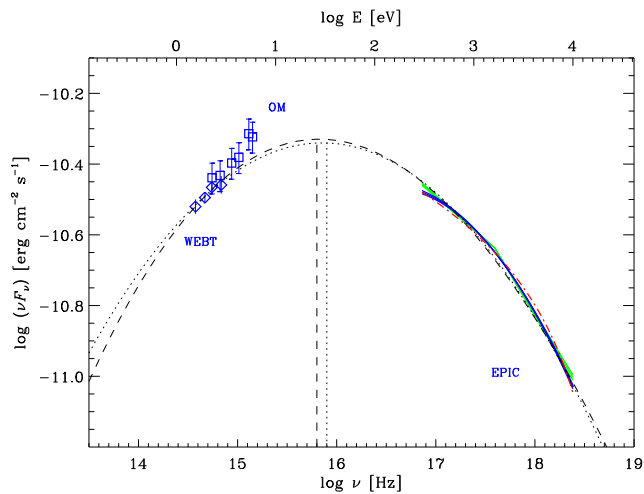}}
   \caption{Spectral energy distribution of PG 1553+113 during the {\em XMM-Newton} pointing of 2013 July 24--25 ($\rm JD=2456498$). Data from the WEBT are displayed as diamonds, while those from {\em XMM-Newton}-OM as squares. The errors on the WEBT data take into account also the uncertainties on the calibration. The EPIC spectra include errors on the free parameters of the log-parabola (blue) and broken power law (green) models.  
The dotted and dashed lines are log-parabolic and log-cubic fits, respectively, suggesting a synchrotron peak at about 26--33 eV. The red dot-dashed line represents the episodic particle acceleration model by \citet{perlman2005}.}
    \label{sed_xmm}
    \end{figure}

The SED corresponding to the {\em XMM-Newton} pointing of 2013 July 24--25 ($\rm JD=2456498.13560$--2456498.52237) is shown in Fig.\ \ref{sed_xmm}. Both the log-parabola (blue) and the broken power-law (green) fits to the EPIC spectra are plotted. The OM spectrum is hard. The simultaneous WEBT data points are from the Crimean Observatory. The $B$ and $V$ ground-based fluxes appear in agreement, within the errors, with those derived from the space observations. 
The dotted and dashed lines represent log-parabolic and log-cubic fits to the data, respectively\footnote{The $F$-test probability is 0.17, which does not clearly favour any of the two models.}.
They highlight that the OM UV$M2$ and UV$W2$ data points are a bit too high to smoothly connect with the X-ray spectrum, whose curvature cannot be fully accounted for.
The fits suggest a synchrotron peak located at $\log \nu_{\rm p} =15.8$--15.9, i.e.\ $\nu_{\rm p}=26$--33 eV.
For comparison, a parabolic fit through the SED including {\em XMM-Newton} observations in 2001, when the source was 1.5 times brighter, gave $\log \nu_{\rm p} =15.7$ \citep{perlman2005}.
In contrast, the log-parabolic curvature of the 2013 X-ray spectrum alone would imply $\log \nu_{\rm p} \sim 16.5$ at $\log \nu F_\nu \sim -10.46$, incompatible with a single synchrotron peak that also includes the UV and optical data.

\section{Discussion}
\label{discussion}
In the previous section we presented broad-band SEDs of PG 1553+113 at different epochs, where UV and X-ray data from either {\em Swift} or {\em XMM-Newton} were available. The SEDs built with {\em Swift} data present a convexity in the optical--UV that makes a smooth connection of the UV with the X-ray spectra hard to trace, especially when adopting a curved model for the X-ray spectrum. This problem does not depend on the recalibration procedure we applied, since standard calibration even worsens the picture. The errors on the UVOT points however are rather large as well as the uncertainty affecting the XRT spectra.

In contrast, the SED obtained with the {\em XMM-Newton} data is very precise and the problem of connecting the UV to the X-ray spectrum is mitigated, but not completely solved because of the curvature inferred from the spectral analysis of the EPIC data. The polynomial fits suggest a synchrotron peak at a higher frequency than in the {\em Swift} case. It is interesting to notice that a similar X-ray spectral shape was found by \citet{perlman2005} while analysing an {\em XMM-Newton} observation of September 2001 ($\alpha=2.33^{+0.02}_{-0.01}, \, \beta=0.13^{+0.02}_{-0.03}$), and an even more pronounced curvature was obtained by \citet{reimer2008} from a {\em Suzaku} observation in July 2006 ($\alpha=2.19\pm 0.01, \, \beta=0.26\pm 0.01$). In both epochs the X-ray flux was about 1.5 times higher than in the {\em XMM-Newton} observation of 2013. Noticeable curvatures were also estimated by \citet{masf08} from {\em Swift} observations in 2005 ($\alpha=2.11$--2.21, $\beta=0.23$--0.36) and especially from a {\em BeppoSAX} pointing in 1998 ($\alpha=2.17\pm 0.07, \, \beta=0.63\pm 0.08$). In all these cases, the log-parabolic fit of the X-ray spectrum gave a synchrotron peak energy of about 0.5--0.7 keV, i.e.\ more than 10 times higher than we found by fitting the SED over a broader energy range.
Physical explanations of the X-ray spectral curvature observed in BL Lac objects were given by \citet{mas04} 
by assuming that the probability for a particle to increase its energy is a decreasing function of the energy itself,
and by \citet{perlman2005} as due to episodic particle acceleration. 
In the latter model particle acceleration occurs with a typical timescale $T$ and the maximum synchrotron energy $E_1$ is reached at time $T$, when the acceleration stops.
The observed photon spectrum is given by:
$$
{dN \over dE} \propto E^{-(s+1)/2} \, {\rm ln} {{E_1^{1/2} \, (E^{1/2}_{\rm max}-E^{1/2})} \over {E^{1/2} \, (E^{1/2}_{\rm max}-E^{1/2}_1)}}, E \le E_1
$$ 
where $E_{\rm max}$ is the maximum energy that would be reached under steady injection conditions.
We show in Fig.\ \ref{sed_xmm} a fit of this model to the 2013 EPIC spectrum, normalized to 0.3 keV and with the choice of
parameters $s=3.0$, $E_1=25 \, \rm keV$ and $E_{\rm max}=100 \, \rm keV$. 
The model gives a fair fit of the EPIC spectrum, though predicts a slightly stronger curvature at high energies, while the data show a more pronounced bending in soft X-rays.

\begin{figure}
   \centering
   \resizebox{\hsize}{!}{\includegraphics{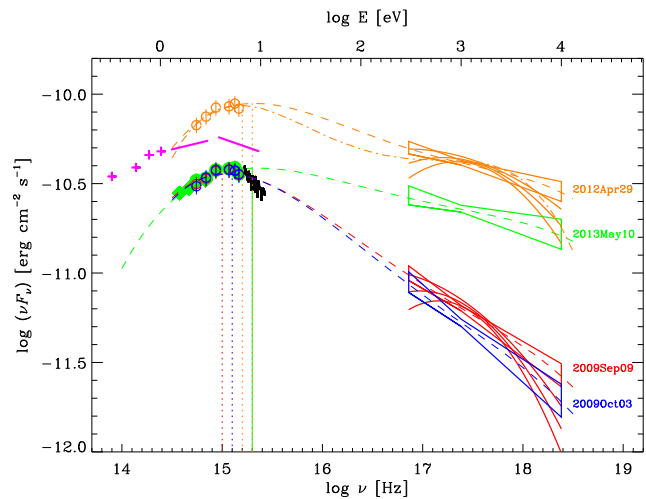}}
   \caption{Spectral energy distributions of PG 1553+113 in different brightness states.
The {\em Swift} data of 2009 September 9, 2009 October 3, 2012 April 29 and 2013 May 10 are shown in red, blue, orange and green, respectively. 
The X-ray spectra are modelled with power laws; in the high and low states also log-parabola fits are shown.
The de-absorbed {\em HST}-COS spectrum of 2009 September 22 is plotted in black and prolongs the {\em Swift}-UVOT spectra of 2009 September 9 and October 3. 
The dashed lines represent fourth-order polynomial fits to the near-IR to power-law X-ray spectra (including the {\em HST}-COS data in 2009); the dot-dashed line is a fit to the optical to log-parabolic X-ray spectrum of 2012 April 29. 
The purple crosses and lines refer to near-IR--UV data acquired in August 1988 close to {\em IUE} observations \citep{falomo1990}.}
    \label{sed_altri}
    \end{figure}

With the aim of clarifying the shape of the source spectra around the synchrotron peak we looked for other data. 
\citet{falomo1990} analysed multiwavelength observations of PG~1553+113 including four pointings by the {\em IUE} in 1982--1988. From their paper we extracted the infrared-to-UV SED corresponding to a high state of the source in August 1988. As can be seen in Fig.\ \ref{sed_altri}, it shows a curvature, suggesting a synchrotron peak around $\log \nu_{\rm p} = 14.9$.

In the same figure we plotted the {\em HST}-COS spectrum acquired on 2009 September 22 and downloaded from the Space Telescope Science Institute archive\footnote{http://mast.stsci.edu}. The spectrum was de-absorbed according to the \citet{car89} law and smoothed\footnote{We also cut the deep absorption line at $\lambda \sim 1210$--1220 \AA\ for sake of clarity.}. The wealth of interstellar and intergalactic medium features were analysed by \citet{danforth2010}, who used them to set constraints on the source redshift (see the Introduction).
The {\em HST} spectrum covers the wavelength range $\sim 1120$--1800 \AA\ and is soft.
It refers to a fainter state of the source than the {\em IUE} spectrum.
{\em Swift} pointed at the source on 2009 September 9 and October 3, i.e.\ 13 days before and 11 days after the {\em HST} observations. 
We analysed the corresponding data following the same procedure described in Sect.\ \ref{swift}.
The results are shown in Fig.\ \ref{sed_altri}, while details on the fit of the XRT spectra are given in Table \ref{xrt} and \ref{logpar}.
While the X-ray flux decreased by $\sim 20\%$ from September 9 to October 3, the optical--UV fluxes are similar in the two epochs.
Hence, if we assume that in that period the UV flux was stable, the composite {\em Swift} plus {\em HST} SED can be considered as indicative of the source spectral shape at that time. A fourth-order polynomial fit shows how the X-ray spectra can smoothly connect with the optical--UV ones, suggesting that the synchrotron peak is at $\log \nu_{\rm p} \approx 15.0$, i.e.\ about 4 eV. This would further confirm the trend of decreasing $\nu_{\rm p}$ with decreasing X-ray flux already seen for the 2013 {\em Swift} observations.
We finally notice that in 2009 the X-ray flux was 3--6 times lower than in 2013 and the X-ray spectrum definitely softer, while the optical--UV level was similar (see Fig.\ \ref{sed_altri}).

Hence, both the {\em IUE} and the {\em HST}-COS spectra confirm the soft {\em Swift}-UVOT spectra, so we can guess that the {\em XMM-Newton}-OM data may suffer calibration problems. Yet, a recalibration of the OM data following the same prescriptions that we used for UVOT does not produce appreciable differences.

The 2013 and 2009 SEDs that we have analysed show faint states of the source.
In order to investigate the SED behaviour at brighter levels, we analysed {\em Swift} data taken on 2012 April 29, at the peak of the 2012 outburst \citep[see also][]{aleksic2015}. 
The results of spectral fitting are reported in Tables \ref{xrt} and \ref{logpar}, while the corresponding SED is displayed in Fig.\ \ref{sed_altri}. The cubic fit suggests a synchrotron peak at $\log \nu_{\rm p} \approx 15.3$, i.e.\ about 8.3 eV.
Because of the higher statistic compared to the other {\em Swift}-XRT observations, we also show the result of a log-parabolic fit on the X-ray spectrum.
This implies a more pronounced curvature than in the {\em XMM-Newton} case and suggests an inflection point around $\log \nu = 16$.

The problem of connecting the optical--UV spectrum with the X-ray one presents no simple solution.
One may wonder whether the choice of Galactic extinction can play a role. We followed \citet{sch98}; had we used the recalibration by \citet{schlafly2011}, which implies a $\sim 20\%$ lower extinction values, we would have obtained a small shift of the optical--UV SEDs downward, by $\sim 0.01$ in the $V$ band, up to $\sim 0.04$ in the $w$2 band. This would slightly improve the match between the {\em XMM-Newton} UV and X-ray spectra, but would worsen it in the {\em Swift} case. In general, it seems hard to find a solution that suits both the {\em Swift} and {\em XMM-Newton} data just playing with the Galactic extinction uncertainties (see also Sect.\ \ref{other}).

Incidentally, we notice that similar connection problems were found for other BL Lac objects, both LBL such as BL Lacertae itself \citep{rav03,rai09} and HBL such as Mkn 421 \citep{mas04} and H 1722+119 (Ahnen et al.\ 2015, in preparation). 
Moreover, inflection points in the infrared portion of the SED have been revealed in 3C 454.3, PKS 1510$-$089 and 3C 279 \citep{weh12,nal12,hay12}. 
All these evidences suggest that a full understanding of blazar emission requires a more complex picture than that is commonly assumed. 

\subsection{Helical jet interpretation}

In the following we will analyse the source SEDs
in an inhomogeneous curved jet scenario, where synchrotron radiation of decreasing frequency is emitted from jet regions at increasing distance from the jet apex, which have different viewing angles and thus different Doppler factors. A helical jet model of this kind was developed by \citet{vil99} and was successfully applied to interpret the observed broad-band flux variability in other BL Lac objects \citep[e.g.][]{rai99,ost04,rai09,rai10}.
According to this model, the jet viewing angle varies along the helical path as
\begin{equation}			\label{costeta}
\cos\theta(z)=\cos\psi\cos\zeta+\sin\psi\sin\zeta\cos(\phi- a z)\,,
\end{equation}
assuming that the helical jet axis lies along the $z$ coordinate of a 3-D reference frame.
The angle $\psi$ is defined by the helix axis and the
line of sight, $\zeta$ is the pitch angle, $\phi$ is the azimuthal difference between the line of sight and the
initial direction of the helical path, and $a$ is the ``curvature", defining the azimuthal angle $\varphi(z)=az$ along the helical path.
Each slice of the jet can radiate, in the plasma rest frame, synchrotron photons from a minimum
frequency $\nu'_{\rm{min}}$ to a maximum one $\nu'_{\rm{max}}$.
The observed flux density has a power-law dependence on the frequency (with spectral index $\alpha_0$) and
a cubic dependence on the Doppler beaming factor $\delta =[\Gamma_{\rm b}(1-\beta\cos\theta)]^{-1}$, where
$\beta$ is the bulk velocity of the emitting plasma
in units of the speed of light, $\Gamma_{\rm b}=(1-\beta^2)^{-1/2}$ the corresponding
bulk Lorentz factor, and $\theta$ is the viewing angle of equation \ref{costeta}.
The variation of the viewing angle along the helical path implies a change of the beaming factor.
As a consequence, the flux at $\nu$ peaks when the part of the jet mostly
contributing to it has minimum $\theta$.

\begin{figure}
   \centering
   \resizebox{\hsize}{!}{\includegraphics{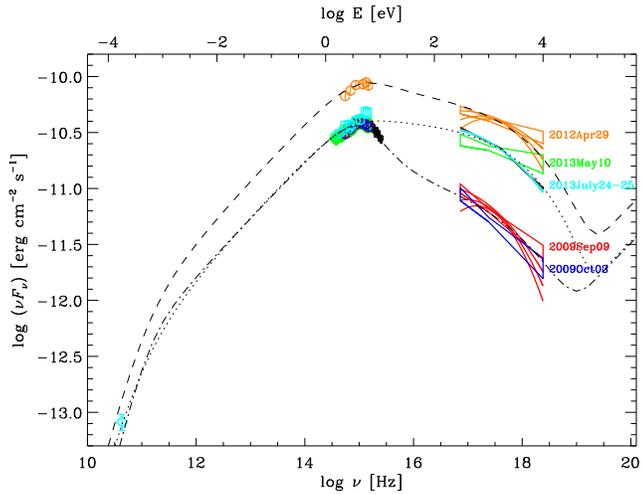}}
   \caption{Spectral energy distributions of PG 1553+113 presented in Figs.\ \ref{sed_xmm} and \ref{sed_altri} fitted with the helical jet model by \citet{vil99}. Three brightness states are presented: the high state of 2012 April 29 (dashed line), the intermediate state of 2013 May/July (dotted line) and the low state of 2009 September/October (dot-dashed line). The fits were obtained by varying only the orientation of the jet, leaving the intrinsic flux unchanged.}
    \label{sed_helssc}
    \end{figure}

In Fig.\ \ref{sed_helssc} we show fits to three different brightness states of the source obtained with this model, where only the three parameters defining the jet orientation ($a$, $\psi$ and $\phi$)
have been changed from one fit to the other, leaving the intrinsic flux unchanged. 
The high state is obtained with $a=45\degr$, $\psi=25.7\degr$ and $\phi=17.5\degr$,
the intermediate state with $a=48\degr$, $\psi=24\degr$ and $\phi=18\degr$, and
the low state with $a=70\degr$, $\psi=24.8\degr$ and $\phi=26.5\degr$.
As for the other parameters introduced above, we fixed $\zeta=30^{\circ}$, $\log \nu'_{\rm min}(0)=16.7$ and $\log \nu'_{\rm max}(0)=19.3$ at jet apex ($z=0$), $\alpha_0=0.5$, and $\Gamma_{\rm b}=10$.
We selected the model parameters in particular to obtain i) an X-ray spectrum with the same curvature as that derived from the {\em XMM-Newton} data, but ii) a UV spectral shape more similar to that given by the {\em Swift}-UVOT data, iii) an inflection point in the SED of the low X-ray state that can reproduce the narrow synchrotron peak marked by the {\em Swift}-UVOT and {\em HST}-COS data, and iv) sufficient flux in the radio band to fit the radio observations at 43 GHz performed close to the {\em XMM-Newton} pointing of 2013 July 24--25.

\begin{figure}
   \centering
   \resizebox{\hsize}{!}{\includegraphics{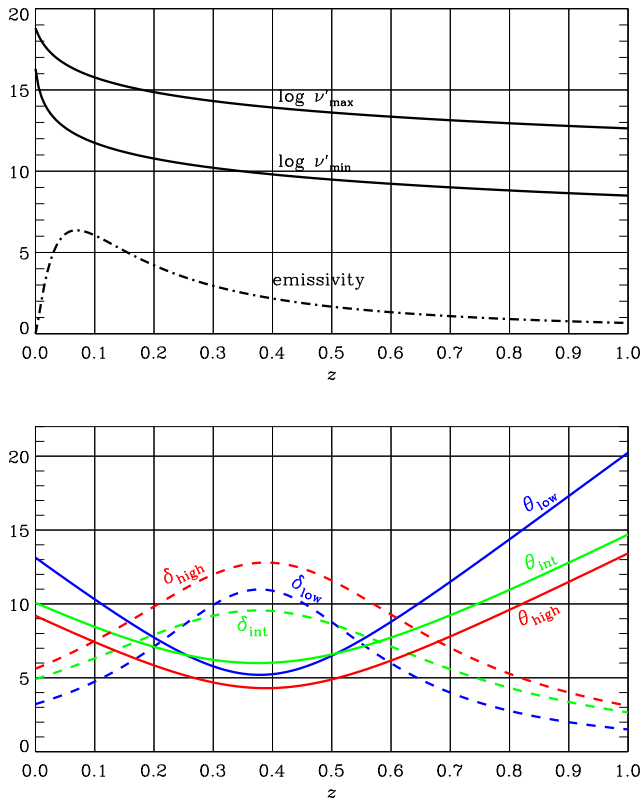}}
   \caption{The helical jet model used to fit the SEDs of Fig.\ \ref{sed_helssc}. Top: the trend of the maximum and minimum emitted frequencies as a function of the distance $z$ along the helical jet axis. The behaviour of the emissivity is also shown (arbitrary units).  Bottom: the viewing angle $\theta$ and Doppler factor $\delta$ along the jet axis for the three fits to the source SEDs presented in Fig.\ \ref{sed_helssc}.}
    \label{andamenti}
    \end{figure}

While we refer to \citet[][and to the other papers mentioned at the beginning of this section]{vil99} for a full mathematical description of the helical jet model, we show in Fig.\ \ref{andamenti} the trends of some physical and geometrical quantities with the distance $z$ along the helix axis. In the top panel one can see that the maximum and minimum emitted frequencies decrease along the jet, while the emissivity first rapidly increases and then slowly decreases\footnote{The laws according to which $\nu'_{\rm min}$ and $\nu'_{\rm max}$ decrease along the jet (see Eqs.\ 4 and 5 in \citealt{vil99}) are defined by: $\log l_1=-2.3$, $\log l_2=-1.8$, $c_1=c_2=3.3$. The trend of emissivity has been slightly modified with respect to Eq.\ 9 in Villata \& Raiteri through the introduction of a further multiplicative factor $(l/l_0)^{(c_0/c_3)}$, where $\log l_0=-1$, $c_0=3$, $c_3=2.3$.}
In the bottom panel we compare the behaviour of the viewing angle $\theta$ and Doppler factor $\delta$
for the three fits to the source SEDs presented in Fig.\ \ref{sed_helssc}.
Close to the jet apex, where the highest frequencies are emitted, the Doppler factor is greater for the high-state fit and lower for the low X-ray state, and this is also true, with different amplitude, along all the jet.
For all cases the maximum beaming occurs around $z=0.38$--0.4, the region from which frequencies in the range $\sim 10^{10}$ to $10^{14} \rm \, Hz$ are emitted.
The peak of the Doppler factor is lower in the intermediate state than in the low state, but $\delta_{\rm int}$ is smoother than $\delta_{\rm low}$, and this explains why the two cases produce the same IR--optical flux. 
Indeed, $\delta_{\rm int}$ soon overcomes $\delta_{\rm low}$, in regions where IR and optical frequencies are still efficiently produced.

In conclusion, the good fits to the source SEDs that the inhomogeneous helical jet model can produce suggest that instabilities developing inside the jet and changing the jet geometry may explain the variations of the SED shape even when the intrinsic jet emission does not change.
We notice that three-dimensional numerical simulations of relativistic
magnetized jets show that kink instabilities are produced, leading to jet wiggling and beam deflection off the main longitudinal axis \citep{mignone2010} or helical jet deformations \citep{moll2008}. These results indicate that variations of the viewing angle of the jet emitting regions are likely to occur.

\subsection{Other possible explanations}
\label{other}
Another possible explanation for the optical--X-ray spectral shape and its variability may be the presence of an extra contribution to the absorption, in addition to the Galactic one, which would imply less steep UV spectra and softer X-ray spectra, eliminating the need for an X-ray spectral curvature.
This is questioned by the {\em XMM-Newton} data, which already present too high UV fluxes and for which the fit of the EPIC spectrum with a power-law model with $N_{\rm H}$ free is statistically worse than the fits with curved models.

An alternative picture is that there are multiple (at least two) emission components, so that the optical--UV and X-ray emissions have a different origin.
This would also explain in particular why the PG 1553+113 X-ray flux varies by about a factor of 4 while the optical--UV level does not change, as can be derived when comparing the state of the source in 2009 September--October with that during the second {\em MAGIC} pointing (see Fig.\ \ref{sed_altri}).

Another possibility is that the energy distribution of the relativistic electrons in the source, $N(\gamma)$, where $\gamma$ is the Lorentz factor of the electrons, has a more complex shape than the usually assumed power-law or log-parabolic one.

To distinguish among the various interpretations, we need more high-quality simultaneous observations in the optical--UV and X-rays, covering different brightness states of the source. This would allow us to better understand the correlation between the emission in the two bands.

\section{Summary and conclusions}

We have presented multiwavelength data acquired during the WEBT campaign organized in 2013 to study the blazar PG 1553+113 during {\em MAGIC} very high energy observations. We have complemented the radio-to-optical data from the WEBT with UV and X-ray observations by {\em Swift} and {\em XMM-Newton} to analyse the synchrotron emission in detail. 
A forthcoming paper will address the higher-energy emission and in particular the results obtained by {\em MAGIC} and {\em Fermi} (Ahnen et al.\ 2015, in preparation).

The source revealed an enigmatic synchrotron behaviour, the connection between the UV and X-ray spectra in the SED appearing more complex than usually assumed. While {\em Swift}-XRT observations have too low statistic to derive detailed information on the X-ray spectral shape, the long exposure by {\em XMM-Newton} in July 2013 confirms that the X-ray spectrum is curved, but probably not as curved as a log-parabolic fit would imply. On the other side, {\em Swift}-UVOT as well as {\em HST}-COS and {\em IUE} data show a steep UV spectrum that hardly connects to the X-ray one.

Among the possible interpretations, we have investigated a scenario where the observed flux variations are due to changes of the viewing angle of the emitting regions in a helical jet, while the intrinsic flux remains constant.
We have obtained good fits to the SEDs built with contemporaneous data in different brightness states by varying only three parameters defining the jet orientation. We are not claiming that all the source variability is due to geometrical effects, but our analysis reveals how important orientation effects are in these beamed sources.
   
We finally note that a periodicity of $720 \pm 60$ d has recently been found when analysing the $\gamma$-ray data of PG 1553+113 \citep{ciprini2014}.
The helical jet model can in principle explain periodicities by assuming that the helix rotates because of the orbital motion in a binary black hole system \citep{vil98a,ost04}. However, a reproduction of both the SED and multiwavelength flux variations requires good data sampling and a fine-tuning of the model parameters.

A further observing effort needs to be spent to fully understand the emission variability mechanisms of this object.

\section*{Acknowledgements}
The data collected by the WEBT collaboration are stored in the WEBT archive; for questions regarding their availability, please contact the WEBT President Massimo Villata ({\tt villata@oato.inaf.it}).
This research has made use of data obtained through the High Energy Astrophysics Science Archive Research Center Online Service, provided by the NASA/Goddard Space Flight Center.
This article is partly based on observations made with the telescopes IAC80 and TCS operated by the Instituto de Astrofisica de Canarias in the Spanish Observatorio del Teide on the island of Tenerife. Most of the observations were taken under the rutinary observation programme. The IAC team acknowledges the support from the group of support astronomers and telescope operators of the Observatorio del Teide.
Based (partly) on data obtained with the STELLA robotic telescopes in Tenerife, an AIP facility jointly operated by AIP and IAC.
This research was (partially) funded by the Italian Ministry for Research and Scuola Normale Superiore.
St.Petersburg University team acknowledges support from Russian RFBR grant 15-02-00949 and St.Petersburg University research grant 6.38.335.2015. AZT-24 observations are made within an agreement between Pulkovo, Rome and Teramo observatories.
The Abastumani team acknowledges financial support of the project FR/638/6-320/12 by the Shota Rustaveli
National Science Foundation under contract 31/77.
Data at Rozhen NAO and Plana SAO were obtained with support by BG051 PO001-3.3.06-0057 project within Human Resources Development Operational Programme 2007-2013.
This research was partially supported by Scientific Research Fund of the Bulgarian Ministry of Education and Sciences under grant DO 02-137 (BIn-13/09). The Skinakas Observatory is a collaborative project of the
University of Crete, the Foundation for Research and Technology -- Hellas, and the Max-Planck-Institut f\"ur Extraterrestrische Physik.
Uzbekistan's authors acknowledge the support from Committee for coordination of science and technology development (Project N.~F2-FA-F027).
G.~D. and O.~V. gratefully acknowledge the observing grant support from the Institute of Astronomy and Rozhen National Astronomical Observatory, Bulgaria Academy of Sciences. This work is a part of the Projects No 176011 (Dynamics and kinematics of celestial bodies and systems), No 176004 (Stellar physics) and No 176021 (Visible and invisible matter in nearby galaxies: theory and observations) supported by the Ministry of Education, Science and Technological Development of the Republic of Serbia. SPM observatory data were obtained through support given by PAPIIT grant​ 111514.
The Mets\"ahovi team acknowledges the support from the Academy of Finland to our observing projects (numbers 212656, 210338, 121148, and others).
Based on observations with the Medicina and Noto telescopes operated by INAF - Istituto di Radioastronomia.

\bibliographystyle{mn2e}

\vspace{1cm}\noindent
{\large \bf AFFILIATIONS}

\vspace{0.5cm}\noindent
{\it
$^{ 1}$INAF, Osservatorio Astrofisico di Torino, via Osservatorio 20, 10025 Pino Torinese, Italy                                                             \\
$^{ 2}$INFN, Sezione di Pisa, Largo Pontecorvo 3, 56127 Pisa, Italy                                                                                          \\
$^{ 3}$Astronomical Institute, St Petersburg State University, 198504 St Petersburg, Russia                                                                  \\
$^{ 4}$Pulkovo Observatory, 196140 St Petersburg, Russia                                                                                                     \\
$^{ 5}$Isaac Newton Institute of Chile, St Petersburg Branch, St Petersburg, Russia                                                                          \\
$^{ 6}$Instituto de Astrofisica de Canarias (IAC), La Laguna, Tenerife, Spain                                                                                \\
$^{ 7}$Departamento de Astrofisica, Universidad de La Laguna, La Laguna, Tenerife, Spain                                                                     \\
$^{ 8}$Institute of Astronomy, Bulgarian Academy of Sciences, 72 Tsarigradsko shosse Blvd, 1784 Sofia, Bulgaria                                              \\
$^{ 9}$Instituto de Astronom\'ia, Universidad Nacional Aut\'onoma de M\'exico, Mexico DF, Mexico                                                             \\
$^{10}$Department of Astronomy, University of Sofia ``St.\ Kliment Ohridski", BG-1164, Sofia, Bulgaria                                                       \\
$^{11}$Crimean Astrophysical Observatory, P/O Nauchny, 298409, Crimea, Russia                                                                                \\
$^{12}$INAF, Osservatorio Astrofisico di Catania, Via S.\ Sofia 78, 95123 Catania, Italy                                                                     \\
$^{13}$Osservatorio Astronomico della Regione Autonoma Valle d''Aosta, Saint-Barth\'elemy, 11020 Nus, Italy                                                  \\
$^{14}$EPT Observatories, Tijarafe, La Palma, Spain,                                                                                                         \\
$^{15}$INAF, TNG Fundaci\'on Galileo Galilei, La Palma, Spain                                                                                                \\
$^{16}$Abastumani Observatory, Mt. Kanobili, 0301 Abastumani, Georgia                                                                                        \\
$^{17}$Astronomical Observatory, Volgina 7, 11060 Belgrade, Serbia                                                                                           \\
$^{18}$INAF, Osservatorio Astronomico di Roma, Monte Porzio Catone, Italy                                                                                    \\
$^{19}$Southern Station of the Sternberg Astronomical Institute, Moscow M.V. Lomonosov State University, P/O Nauchny, 298409, Crimea, Russia                 \\
$^{20}$Ulugh Beg Astronomical Institute, Maidanak Observatory, Uzbekistan                                                                                    \\
$^{21}$INAF, Istituto di Radioastronomia, via Gobetti 101, 40129 Bologna, Italy                                                                              \\
$^{22}$Instituto de Astronom\'ia, Universidad Nacional Aut\'onoma de M\'exico, Ensenada,  BC,  Mexico                                                        \\
$^{23}$Engelhardt Astronomical Observatory, Kazan Federal University, Tatarstan, Russia                                                                      \\
$^{24}$Aalto University Mets\"ahovi Radio Observatory, Mets\"ahovintie 114, 02540 Kylm\"al\"a, Finland                                                       \\
$^{25}$Aalto University Department of Radio Science and Engineering, P.O. BOX 13000, FI-00076 AALTO, Finland.                                                \\
$^{26}$Faculty of Mathematics, University of Belgrade, Studentski trg 16, 11000 Belgrade, Serbia                                                             \\
$^{27}$Instituto Nacional de Astrof\'isica, \'Optica y Electr\'onica, Puebla, Mexico                                                                         \\
$^{28}$Michael Adrian Observatory, Fichtenstrasse 7, 65468 Trebur, Germany                                                                                   \\
$^{29}$INFN, I-35131 Padova, Italy                                                                                                                           \\
$^{30}$INAF-Osservatorio Astronomico di Padova, Vicolo Osservatorio 5, 35122, Padova, Italy                                                                  \\
$^{31}$ETH Zurich, CH-8093 Zurich, Switzerland                                                                                                               \\
$^{32}$ISDC - Science Data Center for Astrophysics, 1290, Versoix, Geneva, Switzerland                                                                       \\
$^{33}$Department of Physics, University of Colorado Denver, CO, USA                                                                                         \\
}
\bsp

\label{lastpage}

\end{document}